\documentclass{article}
\usepackage{amssymb}
\usepackage{amsfonts}
\usepackage{amsmath}
\usepackage{cite}

\begin{document}

\title{Non-standard solutions of relativistic wave equations and decays of
elementary particles}
\author{Andrzej Okni\'{n}ski\thanks{
Email: fizao@tu.kielce.pl} \\
%EndAName
Chair of Mathematics and Physics, Politechnika \'{S}wi\c{e}tokrzyska, \\
Al. 1000-lecia PP 7, 25-314 Kielce, Poland}
\maketitle
\begin{abstract}
We carry out a constructive review of non-standard solutions of relativistic wave equations. 
Such solutions are obtained via splitting of relativistic wave equations written in spinor form. 
All these solutions are also solutions of the Dirac equation and are non-standard because they 
involve higher-order spinors. The main finding is that non-standard solutions describe decaying states. 
\end{abstract}
\section{Introduction}
\label{Introduction}
Recently, we have studied generalized solutions of the Dirac equation which
are also solutions of other relativistic wave equations such as spin $0, 1$
Duffin-Kemmer-Petiau (DKP) equations or spin $1$ Hagen-Hurley equations 
\cite{Okninski2003,Okninski2011,Okninski2012,Okninski2014,Okninski2015a,Okninski2015b,Okninski2016}. 
These solutions have been obtained via splitting of relativistic wave
equations written in spinor form. Such generalized solutions of the Dirac
equation are non-standard because they involve higher-order spinors. In this
work we study solutions and subsolutions of the Dirac equation, DKP
equations, Hagen-Hurley equations as well as spin $\frac{3}{2}$ Fierz-Pauli
equations. Our approach, based on spinor form of relativistic wave equations, 
and involving higher-order spinors (multispinors) can be traced back to 
ideas proposed by Schwinger \cite{Schwinger1966,Schwinger1979}. 

The aim of the present investigation, based on papers  \cite{Okninski2003,Okninski2011,Okninski2012,Okninski2014,Okninski2015a,Okninski2015b,Okninski2016}, 
 is to generalize our study of such non-standard solutions to the case of the Fierz-Pauli equations and 
obtain a broader and unifying picture.

The paper is organized as follows. In the next Section relativistic wave
equations written in standard as well as spinor forms are reviewed. In
Section \ref{NSD} non-standard solutions of relativistic spin equations 
for $s=0,\ \frac{1}{2},\ 1,\ \frac{3}{2}$ are derived and interpreted as 
decaying states. In Section \ref{FB} ideas related to multispinors, Fermi-Bose transformations,  
and supersymmetry are reviewed. In the last Section we discuss our results from broader 
point of view involving theoretical framework discussed in Section \ref{FB}. 
In what follows we use notation and conventions described in \cite{Okninski2003,Okninski2011,Okninski2012}.

\section{Relativistic wave equations}
\label{Relativistic}
\subsection{Spin $\frac{1}{2}$}
\label{Rwe1/2}

The Dirac equation, describing spin $\frac{1}{2}$ elementary particles is:%
\begin{equation}
\gamma ^{\mu }p_{\mu }\Psi =m\Psi ,  \label{Dirac1}
\end{equation}%
where $\gamma ^{\mu }$ are $4\times 4$ matrices fulfilling \cite%
{Dirac1928a,Dirac1928b}:%
\begin{equation}
\gamma ^{\mu }\gamma ^{\nu }+\gamma ^{\mu }\gamma ^{\nu }=2g^{\mu \nu
}I_{4\times 4},  \label{algebra-D}
\end{equation}%
where $g^{\mu \nu }=$ \textrm{diag }$\left( 1,-1,-1,-1\right) $ and $%
I_{4\times 4}$ is a $4\times 4$ unit matrix. In the spinor representation of
the Dirac matrices we have $\Psi =\left( \xi ^{A},\ \eta _{\dot{B}}\right)
^{T}$ \cite{Berestetskii1974}, where $^{T}$ denotes transposition of a
matrix. In spinor formalism the Dirac equation reads: 
\begin{equation}
\left. 
\begin{array}{c}
p^{A\dot{B}}\eta _{\dot{B}}=m\xi ^{A}\smallskip \\ 
p_{A\dot{B}}\xi ^{A}=m\eta _{\dot{B}}%
\end{array}%
\right\} .  \label{Dirac1S}
\end{equation}

\subsection{Spin $0$ and $1$}
\label{Rwe01}

Equations considered in this Subsection, describing spin $0$ and $1$ bosons,
are written as:
\begin{equation}
\beta _{\mu }p^{\mu }\Psi =m\Psi .  \label{DKP-s0,1}
\end{equation}%
Eq. (\ref{DKP-s0,1}) describes a particle with definite mass if $\beta ^{\mu
}$ matrices obey the commutation relations \cite%
{Tzou1957a,Tzou1957b,Okninski1981,Beckers1995a,Beckers1995b}:
\begin{equation}
\sum\nolimits_{\lambda ,\mu ,\nu }\beta ^{\lambda }\beta ^{\mu }\beta ^{\nu
}=\sum\nolimits_{\lambda ,\mu ,\nu }g^{\lambda \mu }\beta ^{\nu },
\label{Tzou}
\end{equation}%
where we sum over all permutations of $\lambda ,\mu ,\nu $.

It was noticed in Ref. \cite{Kemmer1939} that $\beta ^{\mu }$ matrices can
be realized in form:%
\begin{equation}
\beta ^{\mu }=\frac{1}{2}\left( \gamma ^{\mu }\otimes I_{4\times
4}+I_{4\times 4}\otimes \gamma ^{\mu }\right) .  \label{beta}
\end{equation}%
It turns out that such $\beta ^{\mu }$ obey simpler but more restrictive
commutation relations \cite{Duffin1938,Kemmer1939}:%
\begin{equation}
\beta ^{\lambda }\beta ^{\mu }\beta ^{\nu }+\beta ^{\nu }\beta ^{\mu }\beta
^{\lambda }=g^{\lambda \mu }\beta ^{\nu }+g^{\nu \mu }\beta ^{\lambda },
\label{DuffinKemmer}
\end{equation}%
for which Eq.~(\ref{DKP-s0,1}) leads to the Duffin-Kemmer-Petiau (DKP)
theory of spin $0$ and $1$ mesons, see \cite%
{Petiau1936,Duffin1938,Kemmer1939}. This reducible $16$-dimensional
representation (\ref{beta}) of $\beta ^{\mu }$ matrices (denoted as $\mathbf{%
16}$) can be decomposed as $\mathbf{16}=\mathbf{10}\oplus \mathbf{5}\oplus 
\mathbf{1}$. Explicit formulas for the corresponding $10\times 10$ (spin $1$
case) and $5\times 5$ (spin $0$) matrices are given in \cite%
{Kemmer1939,Beckers1995a,Beckers1995b}, while the one-dimensional representation $\mathbf{1%
}$ is trivial, i.e. all $\beta ^{\mu }=0$. In the case of more general Eqs. (\ref%
{Tzou}) there are also other representations of $\beta ^{\mu }$ matrices,
see \cite{Beckers1995a,Beckers1995b} for a review. For example, there are two
representations $\mathbf{7}$\ for which the corresponding $7\times 7$
matrices $\beta ^{\mu }$ yield the Hagen-Hurley equations for spin $1$
bosons \cite{HagenHurley1970,Hurley1971,Hurley1974}.

In the DKP theory of spin $0$ mesons the representation $\mathbf{5}$ is used
(i.e. with $5\times 5$ matrices $\beta ^{\mu }$) for which Eq.\thinspace (%
\ref{DKP-s0,1}) reads:%
\begin{equation}
\left. 
\begin{array}{ccc}
p^{\mu }\psi & = & m\psi ^{\mu } \\ 
p_{\nu }\psi ^{\nu } & = & m\psi%
\end{array}%
\right\} ,  \label{DKP-s0-1}
\end{equation}%
with $\Psi $ in Eq.\thinspace (\ref{DKP-s0,1}) defined as:

\begin{equation}
\Psi =\left( \psi ^{\mu },\psi \right) ^{T}=\left( \psi ^{0},\psi ^{1},\psi
^{2},\psi ^{3},\psi \right) ^{T},  \label{wavef-0}
\end{equation}%
where $^{T}$ denotes transposition.

Within the spinor formalism Eqs. (\ref{DKP-s0-1}) can be written as:%
\begin{equation}
\left. 
\begin{array}{c}
p^{A\dot{B}}\psi =m\psi ^{A\dot{B}} \\ 
p_{A\dot{B}}\psi ^{A\dot{B}}=2m\psi%
\end{array}%
\right\} .  \label{DKP-s0-1S}
\end{equation}%
In the case of spin $1$ mesons the representation $\mathbf{10}$ is used for
which Eq.\thinspace (\ref{DKP-s0,1}) reduces to the Proca equations \cite%
{Proca1936}:%
\begin{equation}
\left. 
\begin{array}{rcc}
p^{\mu }\psi ^{\nu }-p^{\nu }\psi ^{\mu } & = & m\psi ^{\mu \nu } \\ 
p_{\mu }\psi ^{\mu \nu } & = & m\psi ^{\nu }%
\end{array}%
\right\} ,  \label{DKP-s1-1}
\end{equation}%
with $\Psi $ in Eq.\thinspace (\ref{DKP-s0,1}) equal:%
\begin{equation}
\Psi =\left( \psi ^{\mu \nu },\psi ^{\lambda }\right) ^{T}=\left( \psi
^{01},\psi ^{02},\psi ^{03},\psi ^{23},\psi ^{31},\psi ^{12},\psi ^{0},\psi
^{1},\psi ^{2},\psi ^{3}\right) ^{T},  \label{wavef-1}
\end{equation}%
where $\psi ^{\lambda }$ are real and $\psi ^{\mu \nu }$ are purely
imaginary (alternatively, $-\partial ^{\mu }\psi ^{\nu }+\partial ^{\nu
}\psi ^{\mu }=m\psi ^{\mu \nu }$, $\partial _{\mu }\psi ^{\mu \nu }=m\psi
^{\nu }$, where $\psi ^{\lambda }$, $\psi ^{\mu \nu }$ are real). The spin $%
1 $ condition, $p_{\nu }\psi ^{\nu }=0$, follows from the second of
Eqs.\thinspace (\ref{DKP-s1-1}) due to antisymmetry of tensor $\psi ^{\mu
\nu }$. Eqs. (\ref{DKP-s1-1}) can be written in spinor form as \cite%
{Lopuszanski1985,Taub1939}:%
\begin{equation}
\left. 
\begin{array}{c}
p_{A}^{\ \dot{B}}\zeta _{C\dot{B}}+p_{C}^{\,\,\dot{B}}\zeta _{A\dot{B}%
}=2m\eta _{AC} \\ 
p_{\,\ \dot{B}}^{A}\zeta _{A\dot{D}}+p_{\,\,\dot{D}}^{A}\zeta _{A\dot{B}%
}=2m\chi _{\dot{B}\dot{D}} \\ 
p_{A}^{\,\,\dot{C}}\chi _{\dot{B}\dot{C}}+p_{\,\,\dot{B}}^{C}\eta
_{AC}=-2m\zeta _{A\dot{B}}%
\end{array}%
\right\} ,  \label{DKP-s1-1S}
\end{equation}%
with symmetric spinors $\eta _{AC}$, $\chi _{\dot{B}\dot{D}}$.

Spin $1$ bosons can be also described by Hagen-Hurley equations \cite{HagenHurley1970,Hurley1971,Hurley1974}:
\begin{equation}
\left. 
\begin{array}{rcl}
\partial _{\mu }\psi _{\nu }-\partial _{\nu }\psi _{\mu }-i\varepsilon _{\mu
\nu \kappa \lambda }\partial ^{\kappa }\psi ^{\lambda } & = & -m\chi _{\mu
\nu } \\ 
\partial ^{\mu }\chi _{\mu \nu } & = & m\psi _{\nu }%
\end{array}%
\right\},  \label{HH1a}
\end{equation}%
\begin{equation}
\left. 
\begin{array}{rcl}
\partial _{\mu }\varphi _{\nu }-\partial _{\nu }\varphi _{\mu }+i\varepsilon
_{\mu \nu \kappa \lambda }\partial ^{\kappa }\varphi ^{\lambda } & = & 
-m\eta _{\mu \nu } \\ 
\partial ^{\mu }\eta _{\mu \nu } & = & m\varphi _{\nu }%
\end{array}%
\right\}, \label{HH1b}
\end{equation}%
obtained for $7\times 7$ matrices $\beta ^{\mu }$ corresponding to
representation $\mathbf{7}$ ($\chi _{\mu \nu }$, $\eta _{\mu \nu }$ differ
by factor $2$ from tensors defined in \cite{Lopuszanski1978,Lopuszanski1985}%
). In equations (\ref{HH1a}) tensor $\chi _{\mu \nu }$ is
selfdual $\left( \chi _{23}=i\chi _{01}\text{, }\chi _{31}=i\chi _{02}\text{%
, }\chi _{12}=i\chi _{03}\right) $ while $\eta _{\mu \nu }$ in  (\ref{HH1b}) is antiselfdual $%
\left( \eta _{23}=-i\eta _{01}\text{, }\eta _{31}=-i\eta _{02}\text{, }\eta
_{12}=-i\eta _{03}\right) $. Note that since there are two different
equations (\ref{HH1a}), (\ref{HH1b}) there are also two sets of $7\times 7$
matrices $\beta _{\mu }$ fulfilling (\ref{DuffinKemmer}).

More exactly, in the case of Eq. (\ref{HH1a}) we put $\Psi =\left( \chi
_{01},\chi _{02},\chi _{03},\psi _{0},\psi _{1},\psi _{2},\psi _{3}\right)
^{T}$ into Eq. (\ref{DKP-s0,1}) with: 
\begin{equation}
\left. 
\begin{array}{ccl}
\beta _{0} & = & i\left(
-e_{1,5}-e_{2,6}-e_{3,7}+e_{7,3}+e_{6,2}+e_{5,1}\right) \\ 
\beta _{1} & = & -i\left(
-e_{1,4}+ie_{2,7}-ie_{3,6}+ie_{3,6}-ie_{7,2}-e_{4,1}\right) \\ 
\beta _{2} & = & -i\left(
-ie_{1,7}-e_{2,4}+ie_{3,5}-ie_{5,3}-e_{4,2}+ie_{7,1}\right) \\ 
\beta _{3} & = & -i\left(
ie_{1,6}-ie_{2,5}-e_{3,4}-e_{4,3}+ie_{5,2}-ie_{6,1}\right)%
\end{array}%
\right\}, \label{beta-2}
\end{equation}%
where $A\,e_{j,k}$ denotes a non-zero element $A$ of $\beta _{\mu }$ lying
on the intersection of $j$-th row and $k$-th column. These matrices are
unitarily related to the corresponding $\tilde{\beta}_{\mu }$ matrices of
Ref. \cite{Beckers1995a,Beckers1995b} (cf. Eq. (2.24) in part I), i.e. there is such $U$
that $U^{\dagger }\tilde{\beta}_{\mu }U=-\beta _{\mu }$. And another
representation $\mathbf{7}$ is $\beta _{\mu }^{\prime }=-\beta _{\mu }^{\ast
}$ where$~^{\ast }$~is the complex conjugation.

These equations can be cast into spinor form \cite{Dirac1936}:
\begin{equation}
\left. 
\begin{array}{l}
p_{\,\ \dot{B}}^{A}\psi _{A\dot{D}}=m\chi _{\dot{B}\dot{D}},\quad \chi _{%
\dot{B}\dot{D}}=\chi _{\dot{D}\dot{B}} \\ 
p_{A}^{\ \dot{D}}\chi _{\dot{B}\dot{D}}=-m\psi _{A\dot{B}}%
\end{array}%
\right\} ,  \label{HH2a}
\end{equation}%
\begin{equation}
\left. 
\begin{array}{l}
p_{A}^{\ \,\dot{B}}\varphi _{C\dot{B}}=m\eta _{AC},\quad \eta _{AC}=\eta
_{CA} \\ 
p_{\,\ \dot{B}}^{C}\eta _{AC}=-m\varphi _{A\dot{B}}%
\end{array}%
\right\} ,  \label{HH2b}
\end{equation}%
with symmetric spinors $\chi _{\dot{B}\dot{D}}$, $\eta _{AC}$ corresponding
to selfdual and antiselfdual tensors $\chi _{\mu \nu }$, $\varphi _{\mu \nu
} $, respectively \cite{Lopuszanski1978,Lopuszanski1985}. Equations (\ref%
{HH2a}), (\ref{HH2b}) violate parity \cite{Lopuszanski1978,Lopuszanski1985}
and hence could be used in the context of weak interactions.

\subsection{Spin $\frac{3}{2}$}
\label{Rwe3/2}

The Rarita-Schwinger equations for spin $\frac{3}{2}$ particles read \cite%
{Berestetskii1974,Rarita1941}: 
\begin{subequations}
\label{RS}
\begin{eqnarray}
\gamma _{\mu }p^{\mu }\Phi _{\nu } &=&m\Phi _{\nu },  \label{RS1a} \\
\gamma ^{\nu }\Phi _{\nu } &=&0,  \label{RS1b}
\end{eqnarray}
\end{subequations}
where $\Phi _{\nu }$ is a vector-valued spinor and Eq. (\ref{RS1b}) is spin $%
1$ constraint. Strictly speaking, there is also another constraint, $%
\partial ^{\nu }\Phi _{\nu }=0$, which follows from Eqs. (\ref{RS}) \cite%
{Pilling2005}.

The Fierz-Pauli equations, the spinor version of Eqs. (\ref{RS}),\ first
written by Dirac and studied later by Fierz and Pauli \cite%
{Berestetskii1974,Fierz1939,Dirac1936} are:
\begin{subequations}
\label{FP1}
\begin{eqnarray}
&&\left. 
\begin{array}{cc}
p^{A\dot{B}}\eta _{A\dot{B}C} & =0 \\ 
p_{A\dot{B}}\xi ^{A\dot{B}\dot{C}} & =0%
\end{array}%
\right\},  \label{FP1a} \\
&&\left. 
\begin{array}{cc}
p^{D\dot{C}}\eta _{AD}^{\dot{B}} & =m\xi _{A}^{\dot{B}\dot{C}} \\ 
p_{D\dot{C}}\xi _{A}^{\dot{B}\dot{C}} & =m\eta _{AD}^{\dot{B}}%
\end{array}%
\right\},  \label{FP1b}
\end{eqnarray}
\end{subequations}
where $\eta _{AD}^{\dot{B}}=\eta _{DA}^{\dot{B}}$, $\xi _{A}^{\dot{B}\dot{C}%
}=\xi _{A}^{\dot{C}\dot{B}}$ and conditions (\ref{FP1a}) exclude presence of
spin $\frac{1}{2}$\ particles.

The Rarita-Schwinger or Fierz-Pauli equations can be used to describe spin $%
\frac{3}{2}$ particles: a hypothetical gravitino (cf. Section $31.3$ of \cite%
{Weinberg2000}) or baryon resonances (see \cite{Partignani2016} for a zoo of
spin $\frac{3}{2}$ resonances). Existence of spin $\frac{3}{2}$ leptons has been also 
conjectured, see \cite{Ozansoy2016} and references therein. 

\section{Non-standard solutions of relativistic wave equations and decays of
elementary particles}
\label{NSD}
\subsection{Spin $0$}
\label{NSD0}

We start with the spin $0$ DKP equations written in the spinor formalism as (\ref{DKP-s0-1S}). 
Splitting the last of equations (\ref{DKP-s0-1S}), $p^{A\dot{B}}\psi _{A\dot{B%
}}=p^{1\dot{1}}\psi _{1\dot{1}}+p^{1\dot{2}}\psi _{1\dot{2}}+p^{2\dot{1}%
}\psi _{2\dot{1}}+p^{2\dot{2}}\psi _{2\dot{2}}=2m\psi $, we obtain two sets
of equations involving components $\psi _{1\dot{1}},\psi _{1\dot{2}},\psi $
and $\psi _{2\dot{1}},\psi _{2\dot{2}},\psi $, respectively:%
\begin{equation}
\left. 
\begin{array}{rl}
p_{1\dot{1}}\psi  & =m\psi _{1\dot{1}} \\ 
p_{1\dot{2}}\psi  & =m\psi _{1\dot{2}} \\ 
p^{1\dot{1}}\psi _{1\dot{1}}+p^{1\dot{2}}\psi _{1\dot{2}} & =m\psi 
\end{array}%
\right\} ,  \label{const-s0-1}
\end{equation}%
\begin{equation}
\left. 
\begin{array}{rl}
p_{2\dot{1}}\psi  & =m\psi _{2\dot{1}} \\ 
p_{2\dot{2}}\psi  & =m\psi _{2\dot{2}} \\ 
p^{2\dot{1}}\psi _{2\dot{1}}+p^{2\dot{2}}\psi _{2\dot{2}} & =m\psi 
\end{array}%
\right\} ,  \label{const-s0-2}
\end{equation}
each of which describes particle with mass $m$ (we check this substituting $%
\psi _{1\dot{1}}$, $\psi _{1\dot{2}}$ or $\psi _{2\dot{1}}$, $\psi _{2\dot{2}%
}$\ into the third equations). The splitting preserving $p_{\mu }p^{\mu
}\psi =m^{2}\psi $ is possible due to spinor identities, $p_{1\dot{1}}p^{1%
\dot{1}}+p_{2\dot{1}}p^{2\dot{1}}=p_{\mu }p^{\mu }$, $p_{1\dot{2}}p^{1\dot{2}%
}+p_{2\dot{2}}p^{2\dot{2}}=p_{\mu }p^{\mu },$ cf. \cite{Okninski2003}. Thus
equations (\ref{const-s0-1}), (\ref{const-s0-2}) are equivalent to the DKP
equations (\ref{DKP-s0-1S}). We described similar equations in \cite%
{Okninski2011}. From each of equations(\ref{const-s0-1}), (\ref{const-s0-2})
an identity follows: 
\begin{subequations}
\begin{align}
p_{1\dot{2}}\psi _{1\dot{1}}& =p_{1\dot{1}}\psi _{1\dot{2}},
\label{identities0-a} \\
p_{2\dot{2}}\psi _{2\dot{1}}& =p_{2\dot{1}}\psi _{2\dot{2}}.
\label{identities0-b}
\end{align}
\end{subequations}

Equation (\ref{const-s0-1}) and the identity (\ref{identities0-a}), as well
as equation (\ref{const-s0-2}) and the identity (\ref{identities0-b}) can be
written in form of the Dirac equations \cite{Okninski2011}: 
\begin{equation}
\left( 
\begin{array}{cccc}
0 & 0 & p_{1\dot{1}} & p_{2\dot{1}} \\ 
0 & 0 & p_{1\dot{2}} & p_{2\dot{2}} \\ 
p^{1\dot{1}} & p^{1\dot{2}} & 0 & 0 \\ 
p^{2\dot{1}} & p^{2\dot{2}} & 0 & 0%
\end{array}%
\right) \hspace{-0.03in}\left( 
\begin{array}{c}
\psi _{1\dot{1}} \\ 
\psi _{1\dot{2}} \\ 
\psi \\ 
0%
\end{array}%
\right) =m\left( 
\begin{array}{c}
\psi _{1\dot{1}} \\ 
\psi _{1\dot{2}} \\ 
\psi \\ 
0%
\end{array}%
\right) ,  \label{D1}
\end{equation}%
\begin{equation}
\left( 
\begin{array}{cccc}
0 & 0 & p_{1\dot{1}} & p_{2\dot{1}} \\ 
0 & 0 & p_{1\dot{2}} & p_{2\dot{2}} \\ 
p^{1\dot{1}} & p^{1\dot{2}} & 0 & 0 \\ 
p^{2\dot{1}} & p^{2\dot{2}} & 0 & 0%
\end{array}%
\right) \hspace{-0.03in}\left( 
\begin{array}{c}
\psi _{2\dot{1}} \\ 
\psi _{2\dot{2}} \\ 
0 \\ 
\psi%
\end{array}%
\right) =m\left( 
\begin{array}{c}
\psi _{2\dot{1}} \\ 
\psi _{2\dot{2}} \\ 
0 \\ 
\psi%
\end{array}%
\right) ,  \label{D2}
\end{equation}%
respectively, with one zero component. Since in Eqs. (\ref{D1}), (\ref{D2})
there is the same differential operator we can write these equations as a
single Dirac equation. Substituting explicit formulas for the spinors $p^{A%
\dot{B}}$, $p_{A\dot{B}}$ \cite{Okninski2003}, we have \cite{Okninski2015b}:

\begin{equation}
\left( p^{0}\gamma ^{0}-p^{1}\gamma ^{1}-p^{2}\gamma ^{2}-p^{3}\gamma
^{3}\right) \mathbb{A}=m\mathbb{A},  \label{Dirac-s=0}
\end{equation}%
where $\mathbb{A}=\left( 
\begin{array}{cc}
\psi _{1\dot{1}} & \psi _{2\dot{1}} \\ 
\psi _{1\dot{2}} & \psi _{2\dot{2}} \\ 
\psi & 0 \\ 
0 & \psi%
\end{array}%
\right) $ is a generalized (matrix) wavefunction and $\gamma ^{\mu }$
matrices read \cite{Okninski2015b}:%
\begin{equation}
\begin{array}{ll}
\gamma ^{0}=\left( 
\begin{array}{cccc}
0 & 0 & 1 & 0 \\ 
0 & 0 & 0 & 1 \\ 
1 & 0 & 0 & 0 \\ 
0 & 1 & 0 & 0%
\end{array}%
\right) , & \gamma ^{1}=\left( 
\begin{array}{cccc}
0 & 0 & 0 & 1 \\ 
0 & 0 & 1 & 0 \\ 
0 & -1 & 0 & 0 \\ 
-1 & 0 & 0 & 0%
\end{array}%
\right) \medskip , \\ 
\gamma ^{2}=\left( 
\begin{array}{cccc}
0 & 0 & 0 & -i \\ 
0 & 0 & i & 0 \\ 
0 & i & 0 & 0 \\ 
-i & 0 & 0 & 0%
\end{array}%
\right) , & \gamma ^{3}=\left( 
\begin{array}{cccc}
0 & 0 & 1 & 0 \\ 
0 & 0 & 0 & -1 \\ 
-1 & 0 & 0 & 0 \\ 
0 & 1 & 0 & 0%
\end{array}%
\right) ,%
\end{array}
\label{gamma1}
\end{equation}
and we use a shorthand 
$\mathbb{A}=\left( \psi _{A\dot{B}},\ \psi
I_{2\times 2}\right) ^{T}$ where $I_{2\times 2}$ is the $2\times 2$ unit
matrix.
This is the modified spinor representation of matrices $\gamma^\mu$ with $\gamma
^{i}\rightarrow -\gamma ^{i}~\left( i=1,2,3\right) $ and $\Psi =\left( \xi
^{A},\ \eta _{\dot{B}}\right) ^{T}\rightarrow \Psi =\left( \eta _{\dot{B}},\
\xi ^{A}\right) ^{T}$ with respect to \cite{Berestetskii1974}.

\subsection{Spin $1$}
\label{NSD1}

We shall now describe two-step splitting of the spin $1$ DKP equations written in the 
spinor notation as  (\ref{DKP-s1-1S}), see Ref. \cite{Okninski2016}. 

Equations (\ref{DKP-s1-1S}) are split to yield two separate equations for
spinors $\chi _{\dot{B}\dot{D}}$, $\zeta _{A\dot{B}}$ and $\eta _{AC}$, $%
\zeta _{A\dot{B}}$:%
\begin{equation}
\left. 
\begin{array}{l}
p_{A}^{\ \,\dot{B}}\zeta _{C\dot{B}}=m\eta _{AC},\quad \eta _{AC}=\eta _{CA}
\\ 
p_{\,\ \dot{B}}^{C}\eta _{AC}=-m\zeta _{A\dot{B}}%
\end{array}%
\right\} ,  \label{DKP-s1-3a}
\end{equation}%
\begin{equation}
\left. 
\begin{array}{l}
p_{\,\ \dot{B}}^{A}\zeta _{A\dot{D}}=m\chi _{\dot{B}\dot{D}},\quad \chi _{%
\dot{B}\dot{D}}=\chi _{\dot{D}\dot{B}} \\ 
p_{A}^{\ \dot{D}}\chi _{\dot{B}\dot{D}}=-m\zeta _{A\dot{B}}%
\end{array}%
\right\} ,  \label{DKP-s1-3b}
\end{equation}%
respectively \cite{Okninski2003}. This first level of splitting was achieved due to
appropriate spinor identities, see Eq.\thinspace (11) in Ref. \cite%
{Okninski2003}. Indeed, solutions of Eqs.\thinspace (\ref{DKP-s1-3a}), (\ref%
{DKP-s1-3b}) obey the DKP equations (\ref{DKP-s1-1S}). We have thus obtained
the Hagen-Hurley equations (\ref{HH2a}), (\ref{HH2b}) in spinor form with $%
\psi _{A\dot{B}}=\varphi _{A\dot{B}}\equiv \zeta _{A\dot{B}}$ (see also Eqs.
(\ref{HH1a}), (\ref{HH1b}) for the tensor setting).

We shall now split the $s=1$ Hagen-Hurley equations (\ref{DKP-s1-3b}).
Substituting expressions for $p_{A}^{\ \,\dot{B}}$ and $p_{\,\ \dot{B}}^{C}$
into Eqs. (\ref{DKP-s1-3b}), cf. \cite{Okninski2003}, we obtain a system of
eight equations: 
\begin{subequations}
\label{SPLIT}
\begin{eqnarray}
&&\left. 
\begin{array}{rcr}
-\left( p^{1}+ip^{2}\right) \chi _{\dot{1}\dot{1}}-\left( p^{0}-p^{3}\right)
\chi _{\dot{2}\dot{1}} & = & -m\zeta _{1\dot{1}} \\ 
\left( p^{0}+p^{3}\right) \chi _{\dot{1}\dot{1}}+\left( p^{1}-ip^{2}\right)
\chi _{\dot{2}\dot{1}} & = & -m\zeta _{2\dot{1}} \\ 
-\left( p^{1}-ip^{2}\right) \zeta _{1\dot{1}}-\left( p^{0}-p^{3}\right)
\zeta _{2\dot{1}} & = & m\chi _{\dot{1}\dot{1}} \\ 
\left( p^{0}+p^{3}\right) \zeta _{1\dot{1}}+\left( p^{1}+ip^{2}\right) \zeta
_{2\dot{1}} & = & m\chi _{\dot{2}\dot{1}}%
\end{array}%
\right\},  \label{split2a} \\
&&\left. 
\begin{array}{rcr}
-\left( p^{1}+ip^{2}\right) \chi _{\dot{1}\dot{2}}-\left( p^{0}-p^{3}\right)
\chi _{\dot{2}\dot{2}} & = & -m\zeta _{1\dot{2}} \\ 
\left( p^{0}+p^{3}\right) \chi _{\dot{1}\dot{2}}+\left( p^{1}-ip^{2}\right)
\chi _{\dot{2}\dot{2}} & = & -m\zeta _{2\dot{2}} \\ 
-\left( p^{1}-ip^{2}\right) \zeta _{1\dot{2}}-\left( p^{0}-p^{3}\right)
\zeta _{2\dot{2}} & = & m\chi _{\dot{1}\dot{2}} \\ 
\left( p^{0}+p^{3}\right) \zeta _{1\dot{2}}+\left( p^{1}+ip^{2}\right) \zeta
_{2\dot{2}} & = & m\chi _{\dot{2}\dot{2}}%
\end{array}%
\right\},  \label{split2b}
\end{eqnarray}
\end{subequations}
where the condition $\chi _{\dot{B}\dot{D}}=\chi _{\dot{D}\dot{B}}$, is not
imposed and all equations are arranged into two subsets (\ref{split2a}), (\ref{split2b}). 
Note that each of these equations is the Dirac equation with the same set of 
$\gamma^\mu$ matrices \cite{Okninski2015b}. 

We note that solutions of two Dirac equations (\ref{SPLIT}) are non-standard
since they involve higher-order spinors rather than spinors $\xi _{A}$,$\
\eta _{\dot{B}}$. To interpret Eqs. (\ref{SPLIT}) we put: 
\begin{subequations}
\label{SUB}
\begin{eqnarray}
\chi _{\dot{B}\dot{D}}\left( x\right) &=&\eta _{\dot{B}}\left( x\right)
\alpha _{\dot{D}}\left( x\right),  \label{sub1} \\
\zeta _{A\dot{B}}\left( x\right) &=&\xi _{A}\left( x\right) \alpha _{\dot{B}%
}\left( x\right),  \label{sub2}
\end{eqnarray}
\end{subequations}
where $\alpha _{\dot{A}}\left( x\right) $ is the Weyl spinor, describing
massless neutrinos, while $\eta _{\dot{B}}\left( x\right) $, $\xi _{A}\left(
x\right) $ are the Dirac spinors. Although neutrinos are massive  \cite{Donoghue2014} their masses
are very small hence this approximation should not lead to significant errors.

Note that now $\chi _{\dot{1}\dot{2}}\neq \chi _{\dot{2}\dot{1}}$ and,
accordingly, the spin is not determined -- more exactly, the spin is in the $%
0\oplus 1$ space. It means that we consider not real but virtual (off-shell)
bosons \cite{Thomson2013}. This substitution is in the spirit of the method
of fusion of de Broglie \cite{deBroglie1943,Corson1953} (similar ansatz was
used in the $s=0$ case \cite{Okninski2015a}). After the substitution of (\ref%
{SUB}) into Eqs. (\ref{SPLIT}) we obtain two equations:

\begin{equation}
\left. 
\begin{array}{rcl}
-\left( p^{1}+ip^{2}\right) \eta _{\dot{1}}\alpha _{\dot{A}}-\left(
p^{0}-p^{3}\right) \eta _{\dot{2}}\alpha _{\dot{A}} & = & -m\xi _{1}\alpha _{%
\dot{A}} \\ 
\left( p^{0}+p^{3}\right) \eta _{\dot{1}}\alpha _{\dot{A}}+\left(
p^{1}-ip^{2}\right) \eta _{\dot{2}}\alpha _{\dot{A}} & = & -m\xi _{2}\alpha
_{\dot{A}} \\ 
-\left( p^{1}-ip^{2}\right) \xi _{1}\alpha _{\dot{A}}-\left(
p^{0}-p^{3}\right) \xi _{2}\alpha _{\dot{A}} & = & m\eta _{\dot{1}}\alpha _{%
\dot{A}} \\ 
\left( p^{0}+p^{3}\right) \xi _{1}\alpha _{\dot{A}}+\left(
p^{1}+ip^{2}\right) \xi _{2}\alpha _{\dot{A}} & = & m\eta _{\dot{2}}\alpha _{%
\dot{A}}%
\end{array}%
\right\},  \label{split3}
\end{equation}%
where $\dot{A}=\dot{1},\dot{2}$, and, after substituting solution of the
Weyl equation 
\begin{equation}
p^{A\dot{B}}\alpha _{\dot{B}}=0,  \label{Weyl}
\end{equation}%
$\alpha _{\dot{A}}\left( x\right) =\hat{\alpha}_{\dot{A}}e^{ik\cdot x}$, $%
k^{\mu }k_{\mu }=0$, we get a single Dirac -- equation for spinors $\xi
_{A}\left( x\right) $, $\eta _{\dot{B}}\left( x\right) $: 
\begin{equation}
\left. 
\begin{array}{rcl}
-\left( \tilde{p}^{1}+i\tilde{p}^{2}\right) \eta _{\dot{1}}-\left( \tilde{p}%
^{0}-\tilde{p}^{3}\right) \eta _{\dot{2}} & = & -m\xi _{1} \\ 
\left( \tilde{p}^{0}+\tilde{p}^{3}\right) \eta _{\dot{1}}+\left( \tilde{p}%
^{1}-i\tilde{p}^{2}\right) \eta _{\dot{2}} & = & -m\xi _{2} \\ 
-\left( \tilde{p}^{1}-i\tilde{p}^{2}\right) \xi _{1}-\left( \tilde{p}^{0}-%
\tilde{p}^{3}\right) \xi _{2} & = & m\eta _{\dot{1}} \\ 
\left( \tilde{p}^{0}+\tilde{p}^{3}\right) \xi _{1}+\left( \tilde{p}^{1}+i%
\tilde{p}^{2}\right) \xi _{2} & = & m\eta _{\dot{2}}%
\end{array}%
\right\},  \label{Dirac,s=1}
\end{equation}%
with rescaled momentum operators $\tilde{p}^{\mu } =  p^{\mu }+k^{\mu }$.

Indeed, let us consider for example the first term in the first of equations
(\ref{split3}). It can be written as:
\begin{equation}
\begin{array}{l}
-\alpha _{\dot{A}}\left( p^{1}+ip^{2}\right) \eta _{\dot{1}}-\eta _{\dot{1}%
}\left( p^{1}+ip^{2}\right) \alpha _{\dot{A}}=-\alpha _{\dot{A}}\left(
p^{1}+ip^{2}\right) \eta _{\dot{1}\medskip } \\ 
-\eta _{\dot{1}}\left( k^{1}+ik^{2}\right) \alpha _{\dot{A}}=-\alpha _{\dot{A%
}}\left( \tilde{p}^{1}+i\tilde{p}^{2}\right) \eta _{\dot{1}}%
\end{array}
\label{derivation}
\end{equation}
and thus Eqs. (\ref{split3})  reduce to a single Dirac equation (\ref{Dirac,s=1}) for
spinors $\xi _{A}\left( x\right) $, $\eta _{\dot{B}}\left( x\right) $ since
components $\alpha _{\dot{1}}\left( x\right) $, $\alpha _{\dot{2}}\left(
x\right) $ cancel out. Equations (\ref{Weyl}), (\ref{Dirac,s=1}) describe two
spin $\frac{1}{2}$ particles, one massless and another massive, 
with spins coupling to $s=0$ or $s=1$, i.e. $\frac{1}{2}\otimes \frac{1}{2}=0\oplus 1$.

The above description fits decay of a virtual $W^-$ boson into a lepton and antineutrino, for example \cite{Okninski2016}:
\begin{equation}
W^{-}\longrightarrow e+\bar{\nu}_{e}.  
\label{W-decay}
\end{equation}
A good example is provided by the case of a (three-body) mixed beta decay \cite{Krane1988}: 
\begin{equation}
n\left( \uparrow \right) \longrightarrow \left\{ 
\begin{array}{l}
p\left( \downarrow \right) +\left[ e\left( \uparrow \right) \bar{\nu}%
_{e}\left( \uparrow \right) \right] \qquad \text{Gamow-Teller
transition\smallskip } \\ 
p\left( \uparrow \right) +\left[ e\left( \uparrow \right) \bar{\nu}%
_{e}\left( \downarrow \right) \right] \qquad \text{Fermi transition}%
\end{array}%
\right.  \label{mixed-beta}
\end{equation}%
where products of the $W^{-}$ boson decay (see \cite{Partignani2016}) are
shown in square brackets and $\left( \uparrow \right) $ denotes spin $\frac{1%
}{2}$ -- this seems to correspond well to the proposed transition from Eq. (%
\ref{DKP-s1-3b}) to Eqs. (\ref{Weyl}), (\ref{Dirac,s=1}). Indeed, 
we had to assume that the spin was in the $0\oplus 1$ space to describe the decay,  
see remark before Eqs. (\ref{split3}). 
And this agrees well with decay products of the $W^-$ meson with spins coupling to 
$s=1$ or $s=0$ with the spin change absorbed by spin flip of the proton. 

\subsection{Spin $\frac{1}{2}$}
\label{NSD1/2}

The Dirac equation (\ref{Dirac1}) can be written in spinor notation as \cite%
{Berestetskii1974}:
\begin{equation}
\left. 
\begin{array}{r}
p^{1\dot{1}}\eta _{\dot{1}}+p^{1\dot{2}}\eta _{\dot{2}}=m\xi ^{1} \\ 
p^{2\dot{1}}\eta _{\dot{1}}+p^{2\dot{2}}\eta _{\dot{2}}=m\xi ^{2} \\ 
p_{1\dot{1}}\xi ^{1}+p_{2\dot{1}}\xi ^{2}=m\eta _{\dot{1}} \\ 
p_{1\dot{2}}\xi ^{1}+p_{2\dot{2}}\xi ^{2}=m\eta _{\dot{2}}%
\end{array}%
\right\} .  \label{Dirac2}
\end{equation}%
Obviously, due to relations between components of $p^{A\dot{B}}$ and $p_{C%
\dot{D}}$, $p_{1\dot{1}}=p^{2\dot{2}}$, $p_{1\dot{2}}=-p^{2\dot{1}}$, $p_{2%
\dot{1}}=-p^{1\dot{2}}$, $p_{2\dot{2}}=p^{1\dot{1}}$, the equation (\ref%
{Dirac2}) can be rewritten in terms of components of $p^{A\dot{B}}$ only.
Eq. (\ref{Dirac2}) corresponds to (\ref{Dirac1}) in the spinor
representation of $\gamma $ matrices and $\Psi =\left( \xi ^{1},\xi
^{2},\eta _{\dot{1}},\eta _{\dot{2}}\right) ^{T}$.

For $m\neq 0$ we can define new higher-order spinors:
\begin{eqnarray}
p_{1\dot{1}}\xi ^{1} &=&m\psi _{1\dot{1}}^{1},\quad p_{2\dot{1}}\xi
^{2}=m\psi _{2\dot{1}}^{2},  \label{def1} \\
p_{1\dot{2}}\xi ^{1} &=&m\psi _{1\dot{2}}^{1},\quad p_{2\dot{2}}\xi
^{2}=m\psi _{2\dot{2}}^{2},  \label{def2}
\end{eqnarray}%
where we have:
\begin{equation}
\psi _{1\dot{1}}^{1}+\psi _{2\dot{1}}^{2}=\eta _{\dot{1}},\quad \psi _{1\dot{
2}}^{1}+\psi _{2\dot{2}}^{2}=\eta _{\dot{2}}.  \label{defs3,4}
\end{equation}
The Dirac equations (\ref{Dirac2}) can be now written with help of Eqs. (\ref
{def1}), (\ref{def2}) as:
\begin{equation}
\left. 
\begin{array}{rl}
p_{1\dot{1}}\xi ^{1} & =m\psi _{1\dot{1}}^{1} \\ 
p_{1\dot{2}}\xi ^{1} & =m\psi _{1\dot{2}}^{1} \\ 
p_{2\dot{1}}\xi ^{2} & =m\psi _{2\dot{1}}^{2} \\ 
p_{2\dot{2}}\xi ^{2} & =m\psi _{2\dot{2}}^{2} \\ 
p_{2\dot{2}}\left( \psi _{1\dot{1}}^{1}+\psi _{2\dot{1}}^{2}\right) -p_{2%
\dot{1}}\left( \psi _{1\dot{2}}^{1}+\psi _{2\dot{2}}^{2}\right)  & =m\xi ^{1}
\\ 
p_{1\dot{2}}\left( \psi _{1\dot{1}}^{1}+\psi _{2\dot{1}}^{2}\right) +p_{1%
\dot{1}}\left( \psi _{1\dot{2}}^{1}+\psi _{2\dot{2}}^{2}\right)  & =m\xi ^{2}%
\end{array}%
\right\},   \label{Dirac3}
\end{equation}
where components $p_{A\dot{B}}$ are used throughout.

It follows from Eqs. (\ref{Dirac3}) that the following identities hold:%
\begin{eqnarray}
p_{1\dot{2}}\psi _{1\dot{1}}^{1} &=&p_{1\dot{1}}\psi _{1\dot{2}}^{1},
\label{id1a} \\
p_{2\dot{2}}\psi _{2\dot{1}}^{2} &=&p_{2\dot{1}}\psi _{2\dot{2}}^{2}.
\label{id2a}
\end{eqnarray}%
Taking into account the identities (\ref{id1a}), (\ref{id2a}) we can
decouple Eqs. (\ref{Dirac3}) and write them as a system of the following two
equations:
\begin{equation}
\left. 
\begin{array}{rl}
p_{1\dot{1}}\xi ^{1} & =m\psi _{1\dot{1}}^{1} \\ 
p_{1\dot{2}}\xi ^{1} & =m\psi _{1\dot{2}}^{1} \\ 
p_{2\dot{2}}\psi _{1\dot{1}}^{1}-p_{2\dot{1}}\psi _{1\dot{2}}^{1} & =m\xi
^{1}%
\end{array}%
\right\} ,  \label{constituent1}
\end{equation}
\begin{equation}
\left. 
\begin{array}{rl}
p_{2\dot{1}}\xi ^{2} & =m\psi _{2\dot{1}}^{2} \\ 
p_{2\dot{2}}\xi ^{2} & =m\psi _{2\dot{2}}^{2} \\ 
p_{1\dot{2}}\psi _{2\dot{1}}^{2}+p_{1\dot{1}}\psi _{2\dot{2}}^{2} & =m\xi
^{2}%
\end{array}%
\right\} .  \label{constituent2}
\end{equation}
System of equations (\ref{constituent1}), (\ref{constituent2}) is equivalent
to the Dirac equation (\ref{Dirac2}) if the definitions (\ref{defs3,4}) are invoked. 
Due to the identities (\ref{id1a}), (\ref{id2a})
equations (\ref{constituent1}), (\ref{constituent2}) can be cast into
covariant form, cf. Subsection \ref{NSD0}, note however that some components
of spinor $\psi _{B\dot{C}}^{A}$\ are missing.

The problem of missing components of spinor $\psi _{B\dot{C}}^{A}$ is rather
serious because it means that theory is not fully covariant. To solve the
problem in the spirit of Ref. \cite{Okninski2014} we could assume that $\xi
^{A}\left( x\right) =\hat{\alpha}^{A}\chi \left( x\right) $, $\psi _{B\dot{C}%
}^{A}=\hat{\alpha}^{A}\chi _{B\dot{C}}\left( x\right) $ where $\hat{\alpha}_{%
\dot{B}}$ is a constant spinor. In this work we make more general
assumptions: 
\begin{subequations}
\label{neutrinoF}
\begin{eqnarray}
\xi ^{A}\left( x\right)  &=&\alpha ^{A}\left( x\right) \chi \left( x\right) ,
\label{neutrino1F} \\
\psi _{B\dot{C}}^{A}\left( x\right)  &=&\alpha ^{A}\left( x\right) \chi _{B%
\dot{C}}\left( x\right) ,  \label{neutrino2F}
\end{eqnarray}
\end{subequations}
where $\alpha ^{A}\left( x\right) =\hat{\alpha}^{A}e^{ik\cdot x}$, $k^{\mu
}k_{\mu }=0$, is a two-component neutrino spinor, i.e. it fulfills the Weyl
equation, $p_{A\dot{B}}\alpha ^{A}\left( x\right) =0$. Substituting (\ref%
{neutrinoF}) into Eqs. (\ref{constituent1}), (\ref{constituent2}) we get: 
\begin{equation}
\left. 
\begin{array}{rl}
p_{1\dot{1}}\alpha ^{1}\chi  & =m\alpha ^{1}\chi _{1\dot{1}} \\ 
p_{1\dot{2}}\alpha ^{1}\chi  & =m\alpha ^{1}\chi _{1\dot{2}} \\ 
p_{2\dot{2}}\alpha ^{1}\chi _{1\dot{1}}-p_{2\dot{1}}\alpha ^{1}\chi _{1\dot{2%
}} & =m\alpha ^{1}\chi 
\end{array}%
\right\} ,  \label{constituent3}
\end{equation}%
\begin{equation}
\left. 
\begin{array}{rl}
p_{2\dot{1}}\alpha ^{2}\chi  & =m\alpha ^{2}\chi _{2\dot{1}} \\ 
p_{2\dot{2}}\alpha ^{2}\chi  & =m\alpha ^{2}\chi _{2\dot{2}} \\ 
p_{1\dot{2}}\alpha ^{2}\chi _{2\dot{1}}+p_{1\dot{1}}\alpha ^{2}\chi _{2\dot{2%
}} & =m\alpha ^{2}\chi 
\end{array}%
\right\} ,  \label{constituent4}
\end{equation}
and:%
\begin{equation}
\left. 
\begin{array}{rl}
\tilde{p}_{1\dot{1}}\chi  & =m\chi _{1\dot{1}} \\ 
\tilde{p}_{1\dot{2}}\chi  & =m\chi _{1\dot{2}} \\ 
\tilde{p}_{2\dot{2}}\chi _{1\dot{1}}-\tilde{p}_{2\dot{1}}\chi _{1\dot{2}} & 
=m\chi 
\end{array}%
\right\} ,  \label{decoupled1}
\end{equation}%
\begin{equation}
\left. 
\begin{array}{rl}
\tilde{p}_{2\dot{1}}\chi  & =m\chi _{2\dot{1}} \\ 
\tilde{p}_{2\dot{2}}\chi  & =m\chi _{2\dot{2}} \\ 
\tilde{p}_{1\dot{2}}\chi _{2\dot{1}}+\tilde{p}_{1\dot{1}}\chi _{2\dot{2}} & 
=m\chi 
\end{array}%
\right\} ,  \label{decoupled2}
\end{equation}
where $\tilde{p}^{A\dot{B}}=p^{A\dot{B}}+k^{A\dot{B}}$, since components $%
\alpha ^{A}$ cancel out. We note that equations (\ref{decoupled1}), (\ref%
{decoupled2}) are the set of two $3\times 3$ equations equivalent to the spin $0$ DKP 
equation with rescaled momentum operators $\tilde{p}^{A\dot{B}}$. 

We have thus described a transition from the Dirac equation describing a
spin $\frac{1}{2}$ fermion to spin $0$ massive boson and spin $\frac{1}{2}$ Weyl
particle. In our previous paper \cite{Okninski2015a} we have suggested that inverse 
of the described process corresponds to the first stage of the main
channel of pion decay:%
\begin{equation}
\pi ^{-}\longrightarrow \left( \mu ^{-}\,\bar{\nu}_{\mu }\right)
\longrightarrow \mu ^{-}+\bar{\nu}_{\mu }  \label{decayF1}
\end{equation}%
with formation of the intermediate complex $\left( \mu ^{-}\,\bar{\nu}_{\mu
}\right) $. However, the substitutions (\ref{neutrino1F}), (\ref%
{neutrino2F}) suggest another picture. The transformation  
from spin $\frac{1}{2}$ fermion to spin $0$ boson and spin $\frac{1}{2}$ Weyl particle 
agrees well with two secondary channels of decay of the $\tau $ lepton \cite{Partignani2016}: 
\begin{subequations}
\label{taudecay}
\begin{eqnarray}
\tau ^{-} &\longrightarrow &\pi ^{-}+\nu _{\tau }\quad \left( 10.82\
\%\right)  \label{decayF2a} \\
\tau ^{-} &\longrightarrow &K^{-}+\nu _{\tau }\quad \left( 6.96\times
10^{-3}\right)  \label{decayF2b}
\end{eqnarray}
\end{subequations}

\subsection{Spin $\frac{3}{2}$}
\label{NSD3/2}

We start with Eqs. (\ref{FP1}) making the following substitutions: 
\begin{equation}
\left. 
\begin{array}{c}
\eta _{AC}^{\dot{B}}\left( x\right) =\psi _{A}^{\dot{B}}\left( x\right)
\alpha _{C}\left( x\right) \\ 
\xi _{A}^{\dot{B}\dot{C}}\left( x\right) =\psi _{A}^{\dot{B}}\left( x\right)
\beta ^{\dot{C}}\left( x\right)%
\end{array}%
\right\},  \label{substitutions}
\end{equation}%
where bispinor $\left( \alpha _{A},\ \beta ^{\dot{C}}\right)^T $ is a
wavefunction of a spin $\frac{1}{2}$ fermion and $\psi _{A}^{\dot{B}}$
describes a spin $1$ boson. We thus give up symmetry of spinors 
$\eta _{AC}^{\dot{B}}$,  $\xi _{A}^{\dot{B}\dot{C}}$ 
 in undotted and dotted indices, respectively. 

We have from (\ref{FP1}) and (\ref{substitutions}): 
\begin{subequations}
\label{FP2}
\begin{eqnarray}
&&\left. 
\begin{array}{cc}
p^{A\dot{B}}\psi _{A\dot{B}}\,\alpha _{C} & =0 \\ 
p_{A\dot{B}}\psi ^{A\dot{B}}\,\beta ^{\dot{C}} & =0%
\end{array}%
\right\} ,  \label{FP2a} \\
&&\left. 
\begin{array}{cc}
p^{D\dot{C}}\psi _{A}^{\dot{B}}\,\alpha _{D} & =m\psi _{A}^{\dot{B}}\,\beta
^{\dot{C}} \\ 
p_{D\dot{C}}\psi _{A}^{\dot{B}}\,\beta ^{\dot{C}} & =m\psi _{A}^{\dot{B}%
}\,\alpha _{D}%
\end{array}%
\right\} ,  \label{FP2b}
\end{eqnarray}
\end{subequations}
Let us assume that $\psi _{A}^{\dot{B}}\left( x\right) =\hat{\psi}_{A}^{\dot{%
B}}e^{ik\cdot x}$, $k^{\mu }k_{\mu }=\kappa ^{2}$. Moreover, we put $\alpha
_{A}\left( x\right) =\hat{\alpha}_{A}e^{iq\cdot x}$, $\beta ^{\dot{B}}\left(
x\right) =\hat{\beta}^{\dot{B}}e^{iq\cdot x}$, $q^{\mu }q_{\mu }=\lambda ^{2}
$. Then we get from (\ref{FP2}):
\begin{subequations}
\label{FP6}
\begin{eqnarray}
&&\left. 
\begin{array}{cc}
\bar{p}^{A\dot{B}}\psi _{A\dot{B}} & =0 \\ 
\bar{p}_{A\dot{B}}\psi ^{A\dot{B}} & =0%
\end{array}%
\right\} ,  \label{FP6a} \\
&&\left. 
\begin{array}{cc}
\tilde{p}^{D\dot{C}}\alpha _{D} & =m\beta ^{\dot{C}} \\ 
\tilde{p}_{D\dot{C}}\beta ^{\dot{C}} & =m\alpha _{D}%
\end{array}%
\right\} ,  \label{FP6b}
\end{eqnarray}%
\end{subequations}
with rescaled momentum operators $\bar{p}^{A\dot{B}}=p^{A\dot{B}}+q^{A\dot{B}}$, $\tilde{p}^{D\dot{C}%
}=p^{D\dot{C}}+k^{D\dot{C}}$. We have thus obtained conditions (\ref{FP6a}) and the
(modified) Dirac equation for spin $\frac{1}{2}$ fermion (\ref{FP6b}). It
follows that the theory describes a massive spin $\frac{1}{2}$ fermion (with 
$m^{2}=\kappa ^{2}+2q^{\mu }k_{\mu }+\lambda ^{2}$) and a massive spin $1$
boson ($m^2=\lambda^2$). 
Indeed, conditions (\ref{FP1a}) reduce via (\ref{FP6a}) to: 
\begin{equation}
\bar{p}^{A\dot{B}}\psi _{A\dot{B}}=0,  \label{condition-s=1}
\end{equation}%
i.e. $\bar{p}^{\mu }\psi _{\mu }=0$, and this is spin $1$ condition.

It turns out that there are no decays of spin $\frac{3}{2}$ particles into
massive spin $\frac{1}{2}$ fermion and massive spin $1$ boson \cite%
{Partignani2016}. However,  decay of hypothetical excited spin $\frac{3}{2}$ neutrino, 
$\nu ^{\ast }\longrightarrow eW$, corresponding to transition from Eqs. (\ref{FP1}) to (\ref{FP6}), 
was considered \cite{Ozansoy2016}. 
On the other hand, there are decays with massless spin $1$ boson. Therefore we demand that $\lambda =0$. 

Now, transition from Eqs. (\ref{FP1}) to (\ref{FP6}) is 
consistent for $\lambda =0$ with the decay of the spin $\frac{3}{2}$
baryon $\Omega _{c}\left( 2770\right) ^{0}$ into the spin $\frac{1}{2}$
baryon $\Omega _{c}^{0}$ and a photon \cite{Partignani2016}:%
\begin{equation}
\Omega _{c}\left( 2770\right) ^{0}\longrightarrow \Omega _{c}^{0}+\gamma .
\label{Omegadecay}
\end{equation}%
Note that the $\Omega _{c}\left( 2770\right) ^{0}-\Omega _{c}^{0}$ mass
difference is so small that all strong decay channels are excluded \cite%
{Partignani2016}.

\section{Particle decays, Fermi-Bose transformations and supersymmetry}
\label{FB}
We have demonstrated in the preceding Section that non-standard solutions of the Dirac equation, i.e. 
solutions involving higher-order spinors, 
correspond to decaying states of particles, see Eqs. (\ref{W-decay}), (\ref{taudecay}), (\ref{Omegadecay}). 
In all these cases there is a Fermi-Bose transition. Of course, all these transitions are consistent with spin-statistics theorem,  
yet FB transformations occur and deserve a study from a more general, unifying, point of view. 

It was noticed by Schwinger that a formalism of multispinors (higher-order spinors) provides a unified description of 
particles with arbitrary spins and statistics  \cite{Schwinger1966}. It was shown in \cite{Schwinger1979} that this 
approach leads in a natural way to FB transformations and supersymmetry. 

There are many ideas connected with fermion-boson (FB) analogies in the
literature. For example, there is FB equivalence, FB duality, FB
transmutations, to name a few. 
Important step in understanding FB analogy was made by Polyakov who
discovered possibility of fermion-boson transmutation of elementary
excitations of a scalar field interacting with the topological Chern-Simons
term in ($2+1$) dimensions \cite{Polyakov1988}. Recently, the smooth and
controlled evolution from a fermionic Bardeen-Cooper-Schrieffer (BCS)
superfluid state to a molecular Bose-Einstein condensate (BEC)\ has been
realized in ultracold Fermi gases \cite{Zwerger2012}. 
More on these ideas can be found in Refs. 
\cite{Garbaczewski1975,Garbaczewski1985,Simulik2005,Simulik1998,Simulik2011,Krivsky2017,Rodrigues2017}. 

Supersymmetry provides theoretical framework to describe transformations between particles 
of different statistics \cite{Weinberg2000}. Recently, several supersymmetric systems, concerned 
mainly with anyons in 2+1 dimensions \cite
{Jackiw1991,Plyushchay1991a,Plyushchay1991b,Plyushchay2006,Horvathy2010,Horvathy2010a,Horvathy2010b,Horvathy2010c} 
as well as with the 3+1 dimensional Majorana-Dirac-Staunton theory \cite
{Horvathy2008}, unifying fermionic and bosonic fields, have been described. 
Additional information on FB and supersymmetric ideas can be found in \cite{Wilczek1990,Stone1994}. 

\section{Discussion}
\label{Discussion}
We have demonstrated that, in the non-interacting case, there are non-standard solutions 
of  spin $0,\ \frac{1}{2},\ 1,\ \frac{3}{2}$ relativistic equations which are also solutions 
of the Dirac equation. All these solutions are non-standard since they involve higher-order 
spinors. The major finding is that such solutions correspond to decaying states, 
see Eqs. (\ref{W-decay}), (\ref{taudecay}), (\ref{Omegadecay}).  Note that in 
all these transitions bosons are born from fermions or fermions are born from bosons, all FB 
transformations consistent with spin-statistics theorem. 

There is also a larger picture. The multispinor description of particle fields and sources provides a unification of 
all spins and statistics \cite{Schwinger1966,Schwinger1979}. More exactly, the formalism involving higher-order spinors 
allows the treatment of all spins on equal footing and, as was shown by Schwinger, leads to Fermi-Bose transformations and 
supersymmetry. The present work as well as our earlier papers 
\cite{Okninski2003,Okninski2011,Okninski2012,Okninski2014,Okninski2015a,Okninski2015b,Okninski2016}  devoted to FB transformations and based on multispinor formalism, belong to this circle of ideas. 

Note finally, that there are many other decay channels of baryons which have not been explained by our formalism, 
see for example decay of $\Delta(1232)$ resonance where the main decay 
mode is $N\pi$ while the mode $N\gamma$, analogous to (\ref{Omegadecay}), is secondary \cite{Partignani2016}.

\subsection*{Conflict of Interests}

The author declares that there is no conflict of interests regarding the
publication of this paper.

\end{document}